# *Acquisition of time-frequency localized mechanical properties of biofilms and single cells with high spatial resolution*

Enrique A. López-Guerra[1,2], Hongchen Shen[1], Santiago D. Solares[2,*], Danmeng Shuai[1,+]

[1]Department of Civil and Environmental Engineering, The George Washington University, Washington, DC 20052, USA.

[2]Department of Mechanical and Aerospace Engineering, The George Washington University, Washington, DC 20052, USA.

+Email: danmengshuai@gwu.edu

*Email: ssolares@gwu.edu

**Graphical abstract**

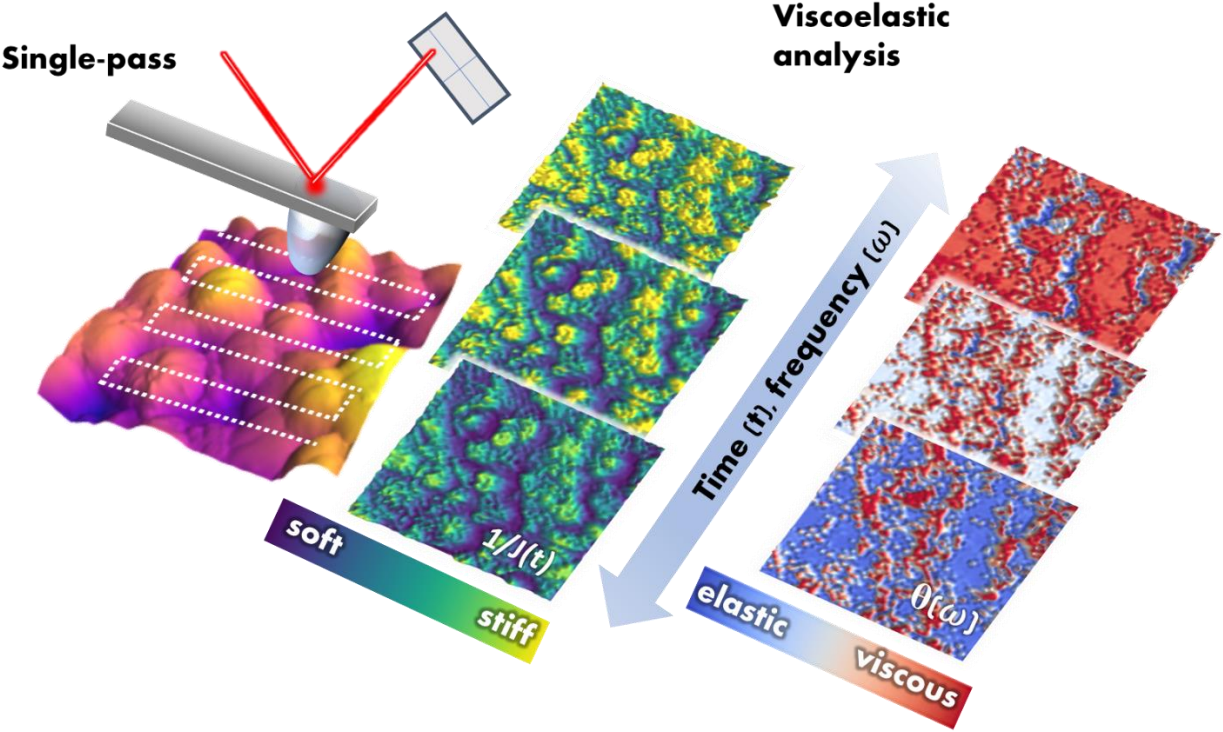

**Abstract**

Biofilms are a cluster of bacteria embedded in extracellular polymeric substances (EPS) that contain a complex composition of polysaccharides, proteins, and extracellular DNA (eDNA). Desirable mechanical properties of the biofilms are critical for their survival, propagation, and dispersal, and the response of mechanical properties to different treatment conditions also sheds light on biofilm control and eradication *in vivo* and on engineering surfaces. However, it is challenging yet important to interrogate mechanical behaviors of biofilms with a high spatial resolution because biofilms are very heterogeneous. Moreover, biofilms are viscoelastic, and their time-dependent mechanical behavior is difficult to capture. Herein, we developed a powerful technique that combines the high spatial resolution of the atomic force microscope (AFM) with a rigorous history-dependent viscoelastic analysis to deliver highly spatial-localized biofilm properties within a wide time-frequency window. By exploiting the use of static force spectroscopy in combination with an appropriate viscoelastic framework, we highlight the intensive amount of time-dependent information experimentally available that has been largely overlooked. It is shown that this technique provides a detailed nanorheological signature of the biofilms even at the single-cell level. We share the computational routines that would allow any user to perform the analysis from experimental raw data. The detailed localization of mechanical properties in space and in time-frequency domain provides insights on the understanding of biofilm stability, cohesiveness, dispersal, and control.

I. **INTRODUCTION**

Biofilms that propagate in human organs and tissues and biomedical devices [1] are one of the leading causes of infectious diseases. Biofilms are also problematic in the food industry by disseminating on food processors, utensils, and packages, ultimately causing foodborne diseases or disease outbreaks[2]. Moreover, biofilms are naturally abundant and undesirable in drinking water distribution pipes where they support the survival and accumulation of pathogens, increasing the risk of waterborne diseases[3]. Biofilms are consortia of microorganisms embedded in a self-produced matrix of extracellular polymeric substances (EPS) that hold the cells together and attach them to surfaces. The EPS are responsible of biofilm mechanical sturdiness by providing biofilms with high cohesion and adhesion stability[4] [5]. Besides the important mechanical stability, EPS are also recognized to enhance antimicrobial and antibiotic resistance of the biofilms[6].

Eradicating biofilms via mechanical failure, either adhesive or cohesive, demands detailed knowledge of biofilm mechanical properties[5]. Specifically, biofilm mechanical properties have been widely recognized to be history-dependent (i.e., viscoelastic). Viscoelasticity refers to the distinct behavior, that some materials such as biofilms have, of simultaneously storing and dissipating energy when deformed. This behavior gives rise to the material appearing to be 'softer or stiffer' and 'less or more dissipative' depending on the rate at which it is probed, fact that adds more complexity to its appropriate analysis. Importantly, this viscoelastic behavior is believed to be responsible for biofilm persistence against mechanical and chemical threats[7-9] (e.g., antibiotics). This belief is supported by observations in polymer physics where it has been shown that adhesion and cohesion failure is a rate dependent phenomenon linked to the polymer viscoelastic properties [10-12]. These observations should be applicable to biofilms where the EPS matrix is regarded as responsible for their outstanding cohesion and adhesion. Despite of its importance, evaluating the mechanics of a highly heterogeneous and complex system is very challenging which has made it impossible to attain a consensus on biofilm mechanical properties[13]. This lack of consensus may have (at least) two distinct roots: *i)* various length scales at which biofilms have been studied (from bulk rheological measurements to micro and nanoscale measurements) [14], and *ii)* whether the studies include or neglect viscoelastic effects.

Within the various length scales, measurements at the nanoscale deserve a special interest due to the high spatial heterogeneity that biofilms display (even at the microscale). Moreover, it is important to understand the underlying mechanisms governing biofilm adhesion and cohesion which are of nanoscale nature (e.g., intermolecular forces between biofilm components). To achieve this goal, the atomic force microscope (AFM) has shown its capability to probe biological systems with nanoscale spatial resolution and high force sensitivity (piconewton) [15-19]. However, the few studies on biofilm nanomechanics available in the literature generally rely on material inversion methods that neglect viscoelastic effects (e.g., Hertzian elastic analysis)[18, 20, 21]. That approach (a first order approximation) has allowed important observations and conclusions, however, to make further advances it is necessary to study in detail the biofilm viscoelastic properties.

The goal of this study is to provide a method that addresses the viscoelastic nature of biofilms while exploiting the high spatial resolution offered by the AFM. The analysis here offered demonstrates the feasibility of localizing mechanical properties spatially and in the time-frequency domain by employing a rigorous framework that considers their history-dependent nature. This technique is of a great interest for understanding in detail the *nanoscale viscoelastic properties* of biofilms, which are understood to be closely related to biofilm stability and cohesiveness, as well as eradication and control. We demonstrate that this method can measure viscoelastic properties with high spatial localization even at the single-bacterium level.

## II. EXPERIMENTAL

**Sample Preparation**

*Staphylococcus epidermidis* (*S. epidermidis*) biofilms were grown over silicon wafers as follows. *S. epidermidis* was cultured in tryptic soy broth (TSB) at 37 °C under mixing conditions overnight and subsequently harvested by centrifugation. Then, bacteria were resuspended in a phosphate buffered saline (PBS) solution and silicon wafers were completely submerged in 2 ml of the bacterial suspension ($OD_{600}$=0.5) in a sterile six-well plate. The system was incubated at 37 °C for 24 h with no stirring to promote bacterial attachment to the silicon wafers. Then, the suspension was evacuated by aspiration, and 10-fold diluted TSB solution was added to submerge the silicon wafers. Last, the wafers were incubated at 37 °C for 3 days with a mixing rate of 80 rpm and daily nutrient replacement of 10-fold diluted TSB solution. Before the AFM experiments, silicon wafers with biofilms were gently rinsed three times by autoclaved ultrapure water and placed in a vacuum dryer overnight at room temperature.

**Atomic force microscopy (AFM) experiments**

Tapping mode and static force spectroscopy AFM techniques were performed to obtain high spatial topographical images and high sensitivity force spectroscopy analysis, respectively. In all experiments presented here we have used commercial AFM cantilevers (Olympus AC200TS R3) with a force constant of approximately 1.1 N/m. We have used a commercial AFM for our measurements (Cypher equipped with an ARC2 controller, Asylum Research).

For morphological characterization of the biofilms we used taping mode AFM. In this popular low invasive technique, the probe is dynamically excited with a sinusoidal signal whose frequency is typically close to the cantilever's first mode resonance frequency [22], and the tapping amplitude is a fraction of the free oscillating amplitude (~50 % in our experiments).

For mechanical characterization we employ static force spectroscopy AFM where the cantilever is approached towards the sample at rates far from the probe's resonance frequency (quasi-static regime) [23, 24]. This mode of operation is very popular for material property calculation because it allows to obtain a force-displacement curve that describes the tip-sample interaction force as a function of the probe position. Consequently, from these force-distance curves material properties are often calculated by employing specific contact mechanic models [17, 25-27]. However, obtaining a force-distance curve requires doing certain assumptions because the AFM does not measure directly force and indentation depth [28]. Instead, it measures values of cantilever deflection, *d(t)*, (in Volts) as a function of the cantilever relative position (also called z-sensor position, *z(t)*). As a result, it is necessary to define unambiguously some reference points (or offsets) during the postprocessing of the raw signals, especially if an automated approach is needed for analyzing multiple force-distance curves as in the present study. This affair is usually not discussed in the literature, although fortunately a couple of reviews on this matter are available [28, 29]. For self-contained purposes, we summarize some key points relevant to our analysis. First, we need to convert the raw signal of deflection (recorded in Volts) to length units through the photodetector sensitivity calibration (with a hard substrate) which yields the deflection history, $d(t)$. Then, we need to capture the point of zero-indentation $(d_0, z_0)$ which in this study was assumed to occur at the value of maximum negative deflection (jump to contact point). Note, that this might not be true for highly adhesive interactions (e.g., Johnson-Kendall-Roberts model JKR[30], Maugis-Dugdale MD[31]) where at the point of contact the indentation can be negative. After defining this point, we can estimate the indentation history, *h(t)*, with the following relationship:

$$h(t) = (z(t) - z_0) - (d(t) - d_0) \qquad (1)$$

In static force spectroscopy, the deflection of the cantilever is directly linked to the tip-sample interaction force. However, prior to multiplying the deflection to the cantilever spring constant ($k_c$) to obtain force, it is needed to subtract a deflection offset ($d_1$) that often appears in the raw signal collected. This offset corresponds to the deflection value in the noncontact region, where tip-sample interaction force is zero. We calculate $d_1$ by averaging the range of $d$ values over the noncontact region (values of deflection far from the jump-to-contact point). With the appropriate offset subtraction, we can then calculate the tip-sample interaction force, *p(t)*:

$$p(t) = k_c(d(t) - d_1) \qquad (2)$$

where the cantilever's stiffness, $k_c$, was calculated through the standard thermal noise method [32].

**Elastic Analysis**

In the first portion of the results section we extract an "apparent stiffness" on the basis of Derjaguin-Muller-Topolov (DMT) theory that combines Hertzian contact mechanics with the inclusion of attractive probe-sample interaction attributed to Van de Waals forces (prominent at the nanoscale) [33]. For the repulsive portion of the interaction, the relationship between tip-sample force, *p*, and sample penetration, *h*, (accessible quantities through the static force spectroscopy experiments) is:

$$p = \frac{4\sqrt{R}}{3} \frac{E}{1-v^2} h^{3/2} - 4\pi R\gamma \qquad (3)$$

where $R, E, v,$ and $\gamma$ are the tip radius, sample Young's modulus, sample Poisson's ratio, and the work of adhesion, respectively. We summarize the parameters accompanying the deformation as: $\beta = \frac{4\sqrt{R}}{3} \frac{E}{1-v^2}$, which is proportional to the Young's modulus. For the measurements in this study we extracted this parameter $\beta$ (instead of the Young's modulus) to avoid doing further assumptions about tip geometry (radius of curvature of the apex, $R$) which is hard to characterize for sharp AFM tips. Afterwards, we normalized the value of stiffness $\beta$ by dividing all measurements by the largest stiffness measurement within the data analyzed. As a result, the values reported range from zero to one (from 'softer' to 'stiffer'). The factor $4\pi R\gamma$ in Equation 3 corresponds to the adhesion force during contact (assumed to be constant in DMT theory) dictated by the work of adhesion between the AFM tip and the biofilm ($\gamma$). This adhesion force was calculated by the minimum tip-sample force located at the jump-to-contact point. Moreover, the AFM tip is assumed to experience no deformation (hard indenter), but instead only the sample is considered to be indented [34, 35].

**Viscoelastic analysis**

The main portion of the viscoelastic analysis (which is the main subject of this study) is presented in the results and discussion section. Here, we show some auxiliary functions that are specifically relevant to our analysis routines. We model the viscoelastic retardance, $U(t)$, with a Prony series to consider the presence of multiple characteristic times [36]. The Prony series can be physically represented in terms of rheological models comprised by springs and dashpots (as we have used in previous studies[37-39]). Specifically, the viscoelastic retardance of the generalized Voigt model (Figure 1) is given by

$$U(t) = J_g + \sum_n J_n/\tau_n \exp(-t/\tau_n) + \{\phi_f\} \qquad (4)$$

where $J_g$ is the *'glassy'* compliance and refers to the material's response at infinitely short time-scales. $\phi_f$ is the steady-state fluidity in the case where the material is regarded as rheodictic (i.e., it can sustain steady-state flow), otherwise if the term is disregarded, the material is considered arrheodictic. $J_n$ and $\tau_n$ refer to the compliance and retardation time of the n[th] Voigt unit in the generalized Voigt model[40].

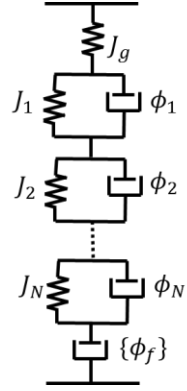

**Figure 1**  The generalized Voigt model consists of an arbitrary large number, N, of springs and dashpots. The elastic compliance of each spring $J_n$ describes how 'soft' the spring is (inverse of stiffness). The fluidity of each dashpot ($\phi_n$) describes how (inversely) viscous the damper is (i.e., how dissipative the dashpot is). This set of springs and dashpots accounts for the simultaneous energy storage and dissipation occurring when a biofilm is deformed. Each spring-dashpot pair in parallel corresponds to a distinct retardation time ($\tau_n = J_n/\phi_n$), i.e., the characteristic time at which rearrangements in the biofilms structure occur when a deformation is imposed. The (strain) response of this model to an impulsive (stress) excitation is given in terms of a Prony series as shown in Equation 4.

In Figure 3 and Figure 4 the inverse of the viscoelastic stiffness was also normalized to be shown as a value ranging from zero to one (from 'softer' to stiffer'). This also obeys practical purposes of avoiding further assumptions about tip geometry (hard to characterize with certainty for a sharp AFM tip). Avoiding assumptions about tip geometry we obtained values of normalized retardance ($U_N(t) = 3/[16\sqrt{R}]U(t)$) from which we obtained values of normalized compliance ($J_N(t) = 3/[16\sqrt{R}]J(t)$). Then in Figure 3 and Figure 4 the normalized values of inverse of stiffness ($1/J_N(t)$) were divided by the maximum value within the data analyzed, resulting in reported values ranging from zero to one.

## III. RESULTS AND DISCUSSION

**3.1 Elastic analysis of biofilms: addressing the spatial heterogeneity**

In this section we show how the atomic force microscope (AFM) can be exploited to obtain mechanical maps of biofilms with nanoscale spatial resolution. We started by employing the common strategy (in AFM studies) of assuming the material to be purely elastic. It is our intention to show that this simplified strategy (although convenient for its simplicity) has significant shortcomings when analyzing a sample that is viscoelastic.

Figure 2 shows the results of performing a mechanical analysis on the biofilm over a scanned image of 5 $\mu m$. Figure 2(a) shows the topographical image acquired with tapping mode AFM where the biofilm structure, comprised by closely packed bacterium cells, is clearly seen. Then we performed force spectroscopy AFM experiments (see Experimental section for details) to interrogate the mechanical properties of the biofilm on a 64×64 pixels grid (higher resolution may be obtained at the expense of a longer experimental time) over the same area shown in Figure 2(a). Specifically, two force spectroscopy maps were acquired at distinct cantilever approach velocities (1.02 and 14.6 $\mu m/s$). Each force map was then postprocessed by assuming that the sample is elastic (see details in Elastic Analysis subsection on Experimental section) and a relative stiffness map was then derived for each distinct approach velocity (Figure 2(b) and (c)). These analyses are quite straightforward and provide with a quick understanding on the mechanical map of the material with high spatial resolution. However, for viscoelastic materials (as the biofilms) the values of apparent stiffness depend on the probing rate, as evidenced by the differences in the maps between Figure 2(b) and (c). Intuitively, viscoelastic materials rearrange stresses when deformed, through processes involving energy dissipation. Therefore, their aparent stiffness depends on the timescale at which they are probed. If the probing timescale is long, the material may rearrange stresses and appear to be 'softer' than if probed at high velocities.

The immediate consequence of the elastic simplification is that inconsistent mechanical maps are obtained when probing a viscoelastic material at different velocities (Figure 2(b) and (c)). This inconsistency causes the stiffness map in Figure 2(b) to display larger bright-color areas, indicating a general apparent stiffer behavior compared to the map in Figure 2(c). This is further evidenced in the summarized statistical analyisis in Figure 2(d) where the histogram for the approach velocity of 14.6 $\mu m/s$ shows significantly larger counts in the higher stiffness range (0.5-1.0) compared to the 1.02 $\mu m/s$ velocity map. The boxplots in Figure 2(d) also support this observation, showing that the distribution of stiffness shifts to higher values for the case of 14.6 $\mu m/s$ approach velocity map. These inconsistencies happen because biofilms have time-dependent mechanical properties, therefore their analysis demand using an appropriate method that considers their viscoelastic behavior.

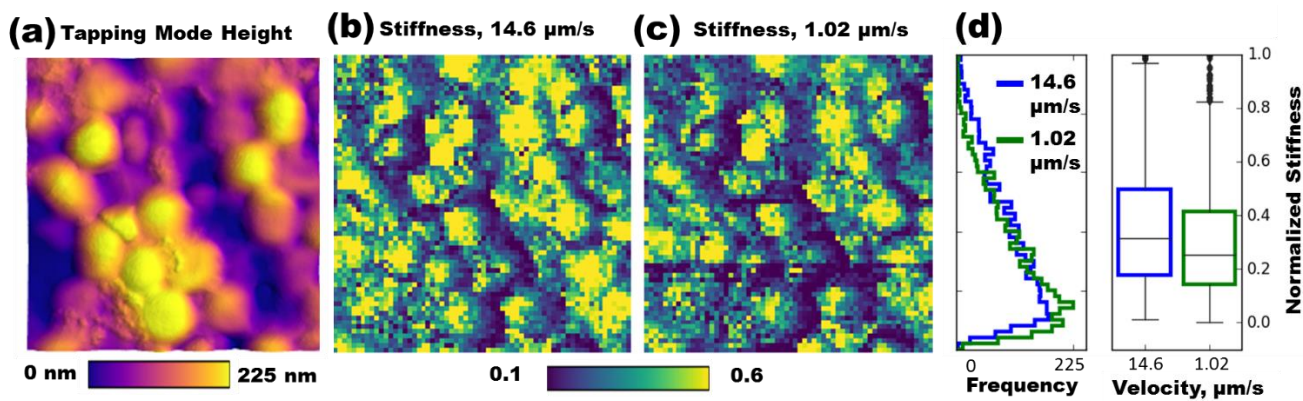

**Figure 2**  Illustration of spatial heterogeneity and velocity dependence of apparent Young's modulus of biofilms. (a) Tapping mode topography of the biofilm over a 5 $\mu m$ size image. (b) and (c) correspond to maps of apparent relative stiffness at two distinct cantilever approach velocities: 14.6 and 1.02 $\mu m/s$, respectively. These stiffness maps correspond approximately to the area of the tapping mode image in (a). The relative stiffness (proportional to apparent Young's modulus) ranges between 0 and 1, from softer to stiffer (see Experimental section for details). (d) Horizontal histogram (left-hand-side) and boxplots (right-hand-side) sharing common vertical axis and legends, for the stiffness data shown in (b) and (c).

**3.2 Viscoelastic analysis: addressing the time and spatial heterogeneity simultaneously**

**3.2.1 Time-frequency localized mechanical maps of biofilms**

As stated in the previous section, the simplified elastic analysis has shortcomings when analyzing viscoelastic materials like biofilms. In this case we should analyze the AFM data in the light of an appropriate theoretical framework that considers the history-dependent nature of biofilms. This distinct nature can be approximately captured by rheological models comprised by (elastic) springs and (viscous) dashpots[39, 40]. The springs reproduce the elastic response of the specimen, whereas the dashpots consider the energy dissipated through the mechanical deformation. The dashpot can be visualized as piston-cylinder device whose mechanical (stress) response is proportional to the (input) strain-rate and the viscosity of the fluid contained in the cylinder. These spring-dashpot models range from very simple sets comprised by one spring and one dashpot (e.g., Maxwell and Kelvin-Voigt) to more sophisticated representations that contain multiple characteristic times (e.g., Generalized Voigt Model as in Figure 1)[39, 40]. The specific model selection obeys practical aspects concerning the time-scale studied, the level of the approximation, the amount of noise in the measurement, etc. However, regardless of the model chosen, viscoelastic materials display general behaviors such as a distinct apparent stiffness and distinct levels of energy dissipation depending on the rate at which they are deformed. A direct consequence of this rate-dependent behavior was discussed in the previous section with respect to the differences in the apparent elastic maps shown in Figure 2(b) and (c). Consequently, viscoelastic materials cannot be described by elastic constants but instead their deformation is captured by time and frequency dependent functions (i.e., the standard viscoelastic responses). For this analysis we chose two distinct standard responses, the loss angle $\theta(\omega)$ and the creep compliance $J(t)$. The creep compliance, $J(t)$, describes how a viscoelastic material deforms in time when a constant stress (force per unit area) is applied [36]. We interpret its inverse, $1/J(t)$, as an intuitive time-localized stiffness (i.e., instantaneous stiffness) for reasons that are mathematically justified in a later subsection.

Figure 3 (a) shows this localized stiffness that we calculate within a finite time window for the same location of the biofilms shown in Figure 2(a). Unlike the elastic analysis of the previous section, in this viscoelastic analysis we are able to capture a time-varying stiffness by exploiting an appropriate mathematical framework that considers the viscoelastic behavior of biofilms. Details on the retrieval of the viscoelastic properties are provided in the following subsection. It is evident that in general the stiffness in the biofilms evolves from a stiffer to a softer behavior (see scale bar in Figure 3(a)). This observation obeys the fact that at short time scales (fast deformations) viscoelastic materials tend to be stiff-elastic, which is known as the glassy response. Intuitively this happens because upon imposition of fast deformations the materials do not have enough time to accommodate the internal stresses and behaves as if they were purely elastic. On the other extreme, at very large time-scales (slow deformations) the material totally rearranges and behaves in a soft-elastic manner, which is known as the rubbery elastic response. The left-hand-side and right-hand-side maps in Figure 3(a) lay in between these two (elastic) extrema in a regime where energy dissipation occurs: the viscoelastic regime. For this analysis we have used the same AFM force spectroscopy raw data used to generate the elastic map in Figure 2(b). Although, in this case instead of retrieving a single map, we obtain a time-localized stiffness that may be plotted with an arbitrary number of frames within the experimental time window. In other words, we can simultaneously localize mechanical properties in space and time. The mathematical details of the analysis are provided in the next subsection.

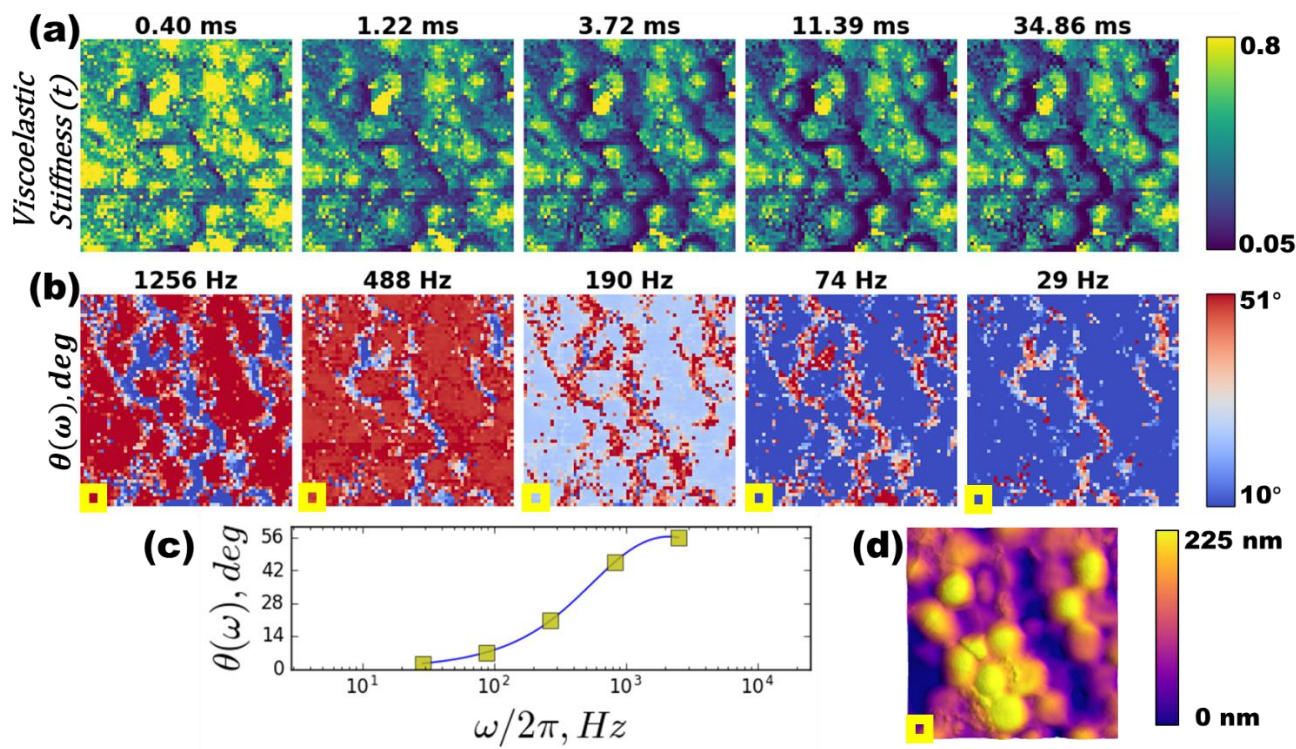

**Figure 3** Representation of the simultaneous spatial and time-frequency localization of mechanical (viscoelastic) properties of biofilms obtained in a single force spectroscopy AFM map. (d) Tapping mode topography of the 5 *μm* image where, approximately, the force spectroscopy map was performed. (a) and (b) highlight the time and frequency localization of mechanical properties, respectively, obtained by the method explained in the next subsection: Mathematical foundation of the viscoelastic analysis. Specifically, we show the localization of mechanical properties in time and frequency in terms of the inverse of viscoelastic compliance, $1/J(t)$, and the loss angle, $\theta(\omega)$, respectively. The time-frequency localization in (a) and (b) is shown from short timescale mechanical behavior (left) to longer timescale behavior (right). The scale bar in (a) shows a value of relative stiffness between 0 and 1 being values closer to 1 an indication of stiffer behavior (details explained in the Experimental section). The units in the scale bar of (b)

is degrees and spans between 0° and 90°, lower values indicating the sample being closely elastic and higher values more viscous. (c) Loss angle $\theta(\omega)$ for the (0,0) spatial pixel, then 5 points equally spaced (in logarithmic frequency scale) are selected to generate frequency-localized maps of loss angle. The calculation of the limits in the time-frequency window is explained in the subsection: Finite time-frequency window of the viscoelastic analysis.

Besides the retrieved time-varying stiffness, the present analysis allows us to retrieve another meaningful viscoelastic property: the loss angle $\theta(\omega)$, whose value spans from zero when the material is purely elastic to ninety when the material is purely viscous. For example, when a biofilm is harmonically deformed with a specific frequency $\omega$ (i.e., sinusoidal input) it dissipates and stores energy with a certain ratio that depends on the value of $\omega$. When the frequency $\omega$ is high (fast deformations) a viscoelastic material tends to its glassy-like behavior, whereas at low excitation frequencies (long time scales) the viscoelastic material tends to behave in a rubber-elastic fashion as previously discussed. We plot $\theta(\omega)$ in Figure 3(b) and it is interesting to observe the complementary information that this quantity offers and how it can further aid in the identification of specific material phases at the surface of the biofilm. In this case, the most abundant phase shows a high viscous behavior at short time scales (high frequencies) while a more elastic-like behavior at long time scales (low frequencies) probably associated to a rubbery response (in consistency with the previous discussion). Here, the frequency window is again finite and defined by the experimental time scale (details in the Experimental Section).

Figure 3(c) shows as an example the loss angle $\theta(\omega)$ retrieved for the (0,0) pixel corresponding to (approximately) the biofilm area marked by the yellow square in Figure 3(d). Once we retrieve this function, we proceed to choose (arbitrarily) five points equally localized in logarithmic scale (marked by the yellow squares in Figure 3(c). By locating these values, we can map them to two-dimensional grids in distinct frames (each frame corresponding to a distinct frequency). By repeating this process for each pixel, we can generate the visualization plotted in Figure 3(b). An analogous process is performed to retrieve the visualization shown in Figure 3(a) showing the instantaneous stiffness. For convenience this process is fully automated, and the analysis is shared in an open-access repository [41].

To demonstrate that this technique can be exploited even at the single-cell scale, we show its applicability in a small scanned area of 500 nm located on the top of a single-bacterium in the biofilm. These results are shown in Figure 4 and the same explanations on Figure 3 apply here. As a result of these analyses we are able to localize viscoelastic properties in time-frequency domain with the high spatial resolution afforded by the AFM. This is a very relevant step forward when compared to the often used Hertzian analysis that allows only to extract a map of apparent stiffness without any time-frequency localization (e.g., Figure 4(d)). By the additional channels that this technique offers, and the time-frequency localization afforded by the viscoelastic analysis, we can better discriminate between components of the materials.

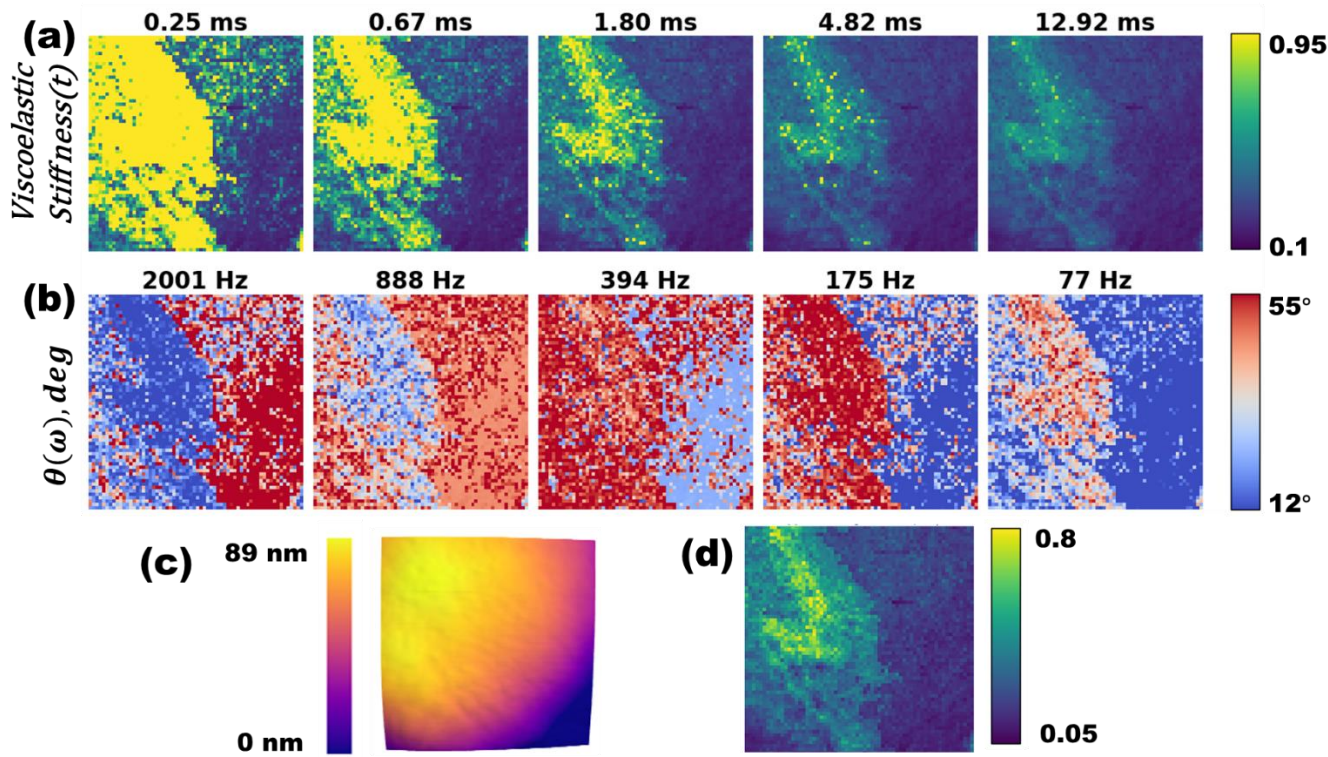

**Figure 4** Same analysis as the one shown in Figure 3 but applied to a 500 nm image size on the top of a single bacterial cell. (c) Tapping mode topography of the top of a bacterial cell. (a) and (b) highlight the time and frequency localization of mechanical properties, respectively. We show the localization of mechanical properties in time and frequency in terms of the inverse of viscoelastic compliance, $1/J(t)$, and the loss angle, $\theta(\omega)$, respectively. The time-frequency localization in (a) and (b) is shown from short timescale mechanical behavior (left) to longer timescale behavior (right). The scale bar in (a) shows a value of relative stiffness between 0 and 1 being ranging from softer to stiffer (details explained in the Experimental section). The units in the scale bar of (b) is degrees and spans between 0° and 90°, lower values indicating the sample being closely elastic and higher values more viscous. (d) 'Apparent stiffness' map when assuming the sample is purely elastic (i.e., using Hertzian analysis), within this framework the time-frequency localization of mechanical properties is not available.

*3.2.2 Mathematical foundation of the viscoelastic analysis*

After the qualitative description, we proceed with a brief mathematical description of the viscoelastic analysis that is used to extract meaningful mechanical information of the biofilm. To gain mechanical information with high spatial resolution, we used a sharp probe with a nanoscale tip that interacted with the sample. Regardless of the rheological model used, when our AFM tip indents a viscoelastic biofilm, the relationship between the loading history, *p(t)*, and indentation, *h(t)*, is given by[39, 42-46]:

$$\frac{16\sqrt{R}}{3} h(t)^{3/2} = \int_0^t U(t-\zeta) p(\zeta) d\zeta \qquad (5)$$

where *R* is the tip radius, *U(t)* the retardance (i.e., the (shear) strain response to a unit (shear) stress impulse [40]), *t* refers to instant time and $\zeta$ is a dummy variable of integration. In this relationship the viscoelastic retardance *U(t)* is convolved with the history of loads *p(t)* applied to the biofilm during the AFM indentation, which underlines the history-dependent nature of the biofilm mechanical model. Equation 5 has assumed the biofilm to be incompressible (Poisson's ratio $\nu = 0.5$), a common assumption in biological models[47, 48]. In a force spectroscopy experiment we can access the indentation and load history (*p(t)*, and *h(t)*) in Equation 5, from which we can (in principle) find a suitable integration kernel, *U(t)*, that satisfies the relationship. Nonetheless, this is a mathematically challenging problem (a first kind Volterra integral equation) that is inherently ill-posed[49, 50]. This issue can be alleviated if assuming we know in advance the general shape of *U(t)*, hence the need of employing specific rheological models (e.g., set of springs and dashpots). In this study we employ the generalized Voigt model[39] (see Figure 1), although the theory outlined here is general and other representations could be employed (e.g., power law models [51, 52], fractional models[53, 54], ladder models[40], Kelvin-Voigt model [16, 55]).

Once we calculate the retardance, *U(t)*, we can easily obtain the so-called creep compliance $J(t)$[40]:

$$J(t) = \int_0^t U(\zeta)d\zeta \tag{6}$$

and the dynamic compliance (also called complex compliance) $J^*(\omega)$[40]:

$$J^*(\omega) = \int_{-\infty}^{\infty} U(t)e^{-i\omega t}dt \tag{7}$$

which has real and imaginary components: $J^*(\omega) = J'(\omega) + iJ''(\omega)$, that correspond to the storage $J'(\omega)$ and loss compliance $J''(\omega)$ respectively. Also, the loss angle, $\theta(\omega)$, can be deduced from the ratio of these quantities:

$$\theta(\omega) = \tan^{-1}\left(\frac{J''(\omega)}{J'(\omega)}\right) \tag{8}$$

This viscoelastic analysis can be performed in a pixel by pixel manner (as illustrated in Figure 5) when the AFM probes specific points in the biofilm while collecting a force spectroscopy map (see Experimental section for details). This map of viscoelastic functions is analogous to the map of stiffness shown in Figure 2(b) and (c), although in this case it is a map of time-dependent or frequency-dependent functions (as shown in Figure 3). Figure 5(b) and (c) show the result of the non-linear square (NLS) optimization process to retrieve the retardance *U(t) for a specific pixel.* In Figure 5(b) the solid lines follow the experimental traces collected by the AFM (i.e., left-hand side of Equation 5) while the dashed lines correspond to the convolution between *U(t)* (obtained through the NLS fitting process), and the experimental load history, *p(t)* (i.e., the right-hand side of Equation 5). The results shown correspond to two distinct cantilever approaching velocities as described by the figure legends. Figure 5(b) shows analogous data as that of Figure 5(a) but in terms of a conventional force-distance curves, a common representation in AFM force spectroscopy.

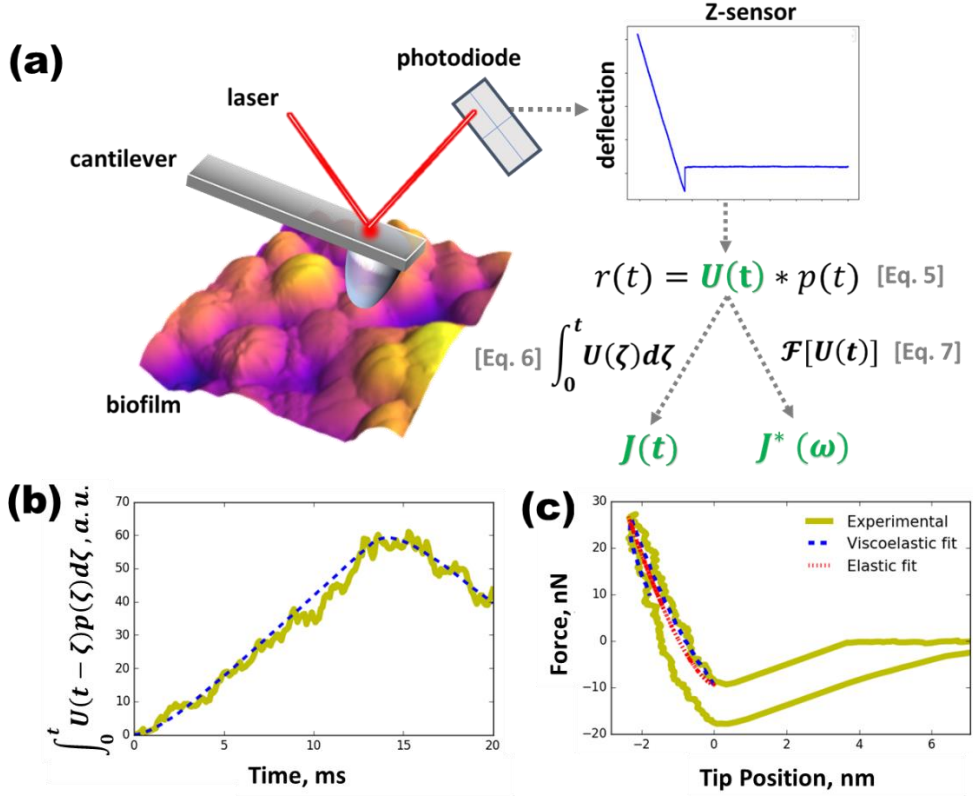

**Figure 5**  Pixel by pixel evaluation of biofilm viscoelastic properties. (a) General Schematics of the experimental procedure and analysis to extract the viscoelastic properties. Briefly, the viscoelastic characterization centers in finding a suitable convolution kernel (the viscoelastic retardance, $U(t)$) in Equation 5 through a non-linear least square optimization (here $r(t)$ summarizes the left hand side of Equation 5). Once the retardance is found, other viscoelastic quantities can be derived from it following Equations 6 - 8. (b) Result of the fitting performed to the force spectroscopy observables in a single pixel. The continuous lines represent the experimental observables (left hand side of Equation 5) while the blue dashed line is the non-linear viscoelastic fits (right hand side of Equation 5). (c) The same results as in (b) but plotted in terms of the more intuitive typical force-distance curve. The elastic fit is also given in the red dashed line

for reference.

### 3.2.3 Why we interpret the inverse of creep compliance (1/J(t)) as a measurement of time-localized stiffness

The relation between load (*p*) and indentation (*h*) described in Equation 5 can be equivalently written in its more popular form in terms of the creep compliance function $J(t)$[42]:

$$\frac{16\sqrt{R}}{3}h(t)^{3/2} = \int_0^t J(t-\zeta)\frac{dp(\zeta)}{d\zeta}d\zeta \qquad (9)$$

During static force spectroscopy AFM, the tip-sample interaction force (load) approximately grows linearly in time: $p \approx \dot{p}\,t$ [25, 26]. Which reduces the previous equations to [39]:

$$\frac{16\sqrt{R}}{3}h(t)^{3/2} = \dot{p}\left[\int_0^t J(\zeta)d\zeta\right] = \dot{p}[\chi(t)] \qquad (10)$$

where the term in brackets, $\chi(t)$, refers to the sample's fluidity, the strain response to a unit stress ramp $(\sigma(t) = \dot{\sigma}\,t)$[40]. Differentiating with respect to time yields:

$$\frac{16\sqrt{R}}{3}\frac{dh(t)^{3/2}}{dt} = \dot{p}J(t) \qquad (11)$$

Now, substituting $\dot{p} = \frac{dp}{dt}$ and rearranging we obtain:

$$\frac{dp}{dh(t)^{3/2}} = \frac{16\sqrt{R}}{3}\frac{1}{J(t)} \qquad (12)$$

We may compare the above with the elastic case (parabolic hard indenter penetrating an elastic surface) [34, 35]

$$\frac{dp}{dh^{3/2}} = \frac{16\sqrt{R}}{3}G \qquad (13)$$

where $G$ is the elastic shear modulus, and the material has been assumed to be incompressible (Poisson's ratio, $\nu = 0.5$). The viscoelastic relationship only differs with the elastic one by the factor $\frac{1}{J(t)}$ in place of $G$. This interesting observation demonstrates that the inverse of the shear creep compliance ($1/J(t)$) is the viscoelastic analog to the elastic shear modulus in a force spectroscopy experiment [56], therefore we use it as a measurement of time-localized (viscoelastic) stiffness in Figure 3 and Figure 4. Recall that shear modulus (*G*) and Young's modulus (*E*) are proportional: $E = 2G(1 + v)$.

*3.2.4 Finite time-frequency window of the viscoelastic analysis*

In this section we discuss some technical yet critical details about the limits in time and frequency where the viscoelastic functions retrieved remain valid. For example, it is desirable to know how dissipative the biofilm is (value of *θ(ω)*) for a wide range of harmonic frequencies *ω*, to know how it will respond to a wide variety of mechanical stimuli. However, our experiment has a limited time-frequency window where our viscoelastic functions calculated are meaningful. In Figure 3 and Figure 4 we have clearly laid out these specific time-frequency windows where the calculations are accurate (from the first to fifth frame), however the justification and calculation of these windows have not been explained yet.

To clearly understand this point it is important to focus our attention to Equation 5 that contains the viscoelastic source function (the retardance, *U(t)*) from which the other functions are derived (Equations 6 - 8). This integration kernel *U(t)* in Equation 5 is the system's response of the material to an impulsive excitation. Any linear theory regards this function as highly desirable since it fully characterizes the system [57, 58]. Therefore, it is tempting to believe that by obtaining a suitable approximation of *U(t)* (by means of Equation 5) would give us information about the material's viscoelastic properties in the whole time and frequency domains. This concept of course is flawed because we only have a finite time resolution of the retrieved *U(t)*. Specifically, we can only access (experimentally) this quantity with a certain degree of certainty within a specific 'region of interest' dictated by the experimental timescale. This 'region of interest' is bound by the limits in which the experimental observables (in this case load and displacement) are above certain level of signal to noise ratio (SNR), i.e., the regions where the quantities deliver meaningful information. For example, if we probe a material at three different speeds (see Figure 6(a)) the retrieved *U(t)* will have different 'regions of interest' for each approach velocity. Only within this 'region of interest' is correct to describe the derived mechanical properties. This concept is illustrated in Figure 6(c) and Figure 6(d) where we have delimited with solid lines the regions where the calculations are appropriate. In the 'signal to noise ratio' (SNR) calculation we made a posteriori estimation, assuming that noise is of stochastic nature (details can be found in Czesla et al.[59]). This assumption is compatible with the type of noise arising from thermal fluctuations of the AFM cantilever[60]. Once noise is calculated, we calculated the minimum time when the deflection signal surpasses the noise level by 10 times (SNR=10) and defined this as the lower limit of the time window ($t_{min}$). The upper limit in the time window ($t_{max}$) is given by the total experimental time during tip-sample contact in the force spectroscopy experiment. The lower limit in the frequency window is the inverse of the total experimental time ($f_{min} = 1/t_{max}$). The upper limit of the frequency window is the Nyquist frequency ($f_{max} = 1/2\, f_s$),[61] where $f_s$ is the sampling frequency, which for the viscoelastic retardance *U(t)* is $1/t_{min}$, thus $f_{max} = 1/[2\, t_{min}]$.

With the dashed lines we highlight the dangers of extrapolating the results to frequency windows beyond the 'region of interest'. In other words, we underline that certain model parameters are only appropriate within a time-frequency window. For this reason, we avoid the practice of reporting specific rheological parameter values and instead focus on discussing standard viscoelastic response (e.g., $J(t), \theta(\omega)$) and the time-frequency regions where they are valid.

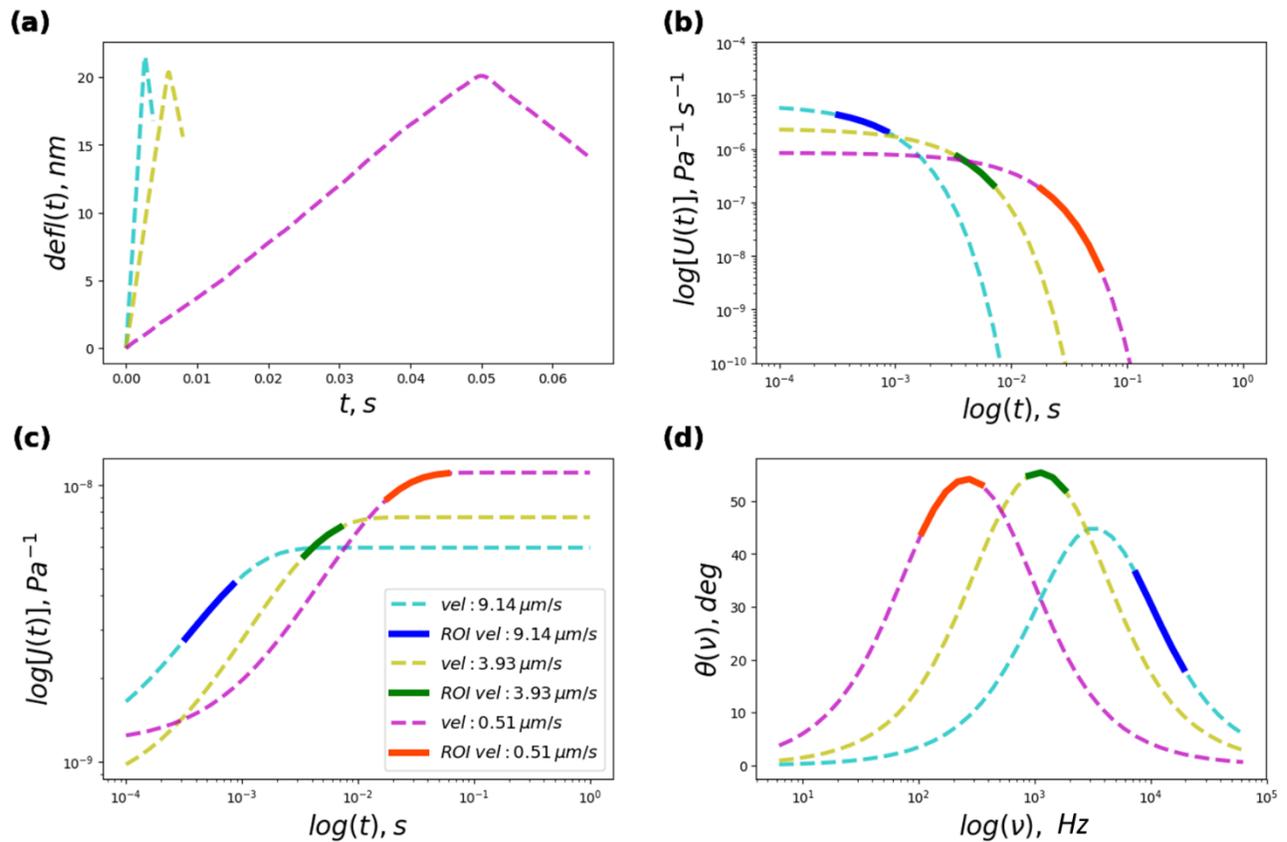

**Figure 6**   Results of viscoelastic analysis related to force spectroscopy experiments performed at three different velocities (as indicated in the figure legends) over approximately the same area (single-pixel) of the biofilm. This figure illustrates the complementary viscoelastic information gained when probing at different speeds. (a) Load histories for three different force spectroscopy experiments. (b)-(d) Calculated viscoelastic quantities: retardance, creep compliance, and loss angle, respectively, with their 'regions of interested (ROI)' delimited within a finite time-frequency window. These viscoelastic quantities were calculated with Equations 5, 6, and 8, respectively. The figure legends shown in (c) are applicable for all subplots. Details on the estimation of the time-frequency window's limits are provided in the main text.

*3.2.4 Implications of the viscoelastic measurements on biofilm removal*

The viscoelastic properties of biofilms, especially of the EPS matrix is believed to confer high structural stability to the system. Therefore, a detail understanding of the viscoelastic signature is highly desirable. Specifically, it is believed that EPS closely relates to the high cohesiveness of the biofilms. Mechanical properties such as shear modulus has been proved to be directly linked to cohesiveness in the context of synthetic polymers. It is logical to extrapolate this observation to biofilms where biopolymers are the main component of the EPS matrix. As it has been discussed in results of Figure 3 and Figure 4, the shear modulus of biofilms is time-dependent (viscoelastic), therefore its apparent value depends on the time scale of the deformation. For example, the time-dependent shear modulus may display a stiff-elastic behavior at fast deformations or soft-elastic behavior at slow deformations. This concept is further illustrated in Figure 7.

For this reason, a viscoelastic framework as the one discussed here is crucial to have a better picture of how biofilms behave at different rates of deformation. This thorough knowledge has direct implications on understanding and envisioning effective techniques for biofilm removal. For example, knowing at which deformation rates the material is less dissipative can give effective guidelines for mechanical removal through time-dependent inputs that can effectively transfer the energy to cohesive fracture. This would also help to avoid deformation rates at which the mechanical input for removal would be wasted through energy dissipation (i.e., viscoelastic dissipation).

The limited time-frequency window in which the biofilm's mechanical properties is evaluated can be conveniently expanded within the framework given in this study (as discussed in the previous subsection, see Figure 6) to higher frequencies up to the kHz regime. This is achieved with the appropriate selection of experimental parameters (e.g., probing velocity, cantilever resonance frequency), which results in obtaining the mechanical response of the biofilm in the ultrasonic regime.

This has direct implications in the context of treating biofilm infections in hospital patients as it has been shown that antimicrobial activity can be enhanced when ultrasound waves are propagated to the specimen. In other words, knowing specifically the nanomechanical response of the biofilm in the ultrasonic regime can provide better understanding on specific frequencies at which the biofilm can become more susceptible to chemical stimuli (e.g., antimicrobial and antibiotic activities). To summarize, the analysis technique presented here allows getting a very thorough viscoelastic picture with high spatial resolution, having explicit practical implications regarding effective mechanisms for biofilm eradication.

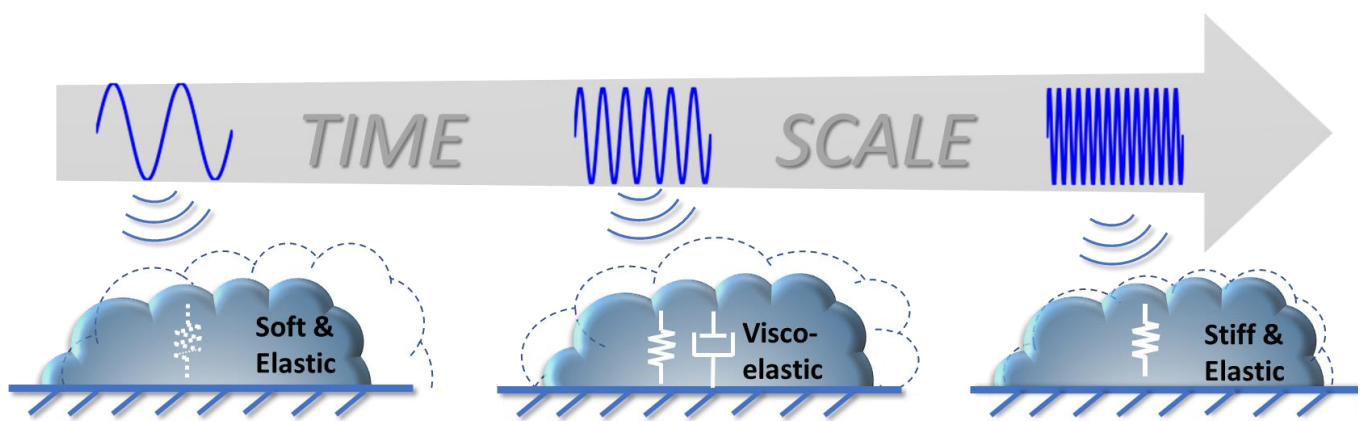

**Figure 7**  Illustrative concept of the viscoelastic properties of biofilms. The conceptual sketch shows that the mechanical behavior of biofilms depends on the timescale of the deformation, showing that at certain deformation rates they may display a *near elastic* behavior whereas at other rates they may be more *dissipative*.

## IV. CONCLUSION

We have presented a powerful technique to obtain viscoelastic properties of nanoscale materials in a localized manner. We have shown that this technique is especially beneficial for highly heterogeneous systems as biofilms. In this powerful technique the mechanical properties are localized in the time-frequency domain giving specific information on how the material behaves at different deformation time scales. The amount of information obtained is very intensive when compared to the often used Hertzian analysis that neglects viscoelastic effects. It has been shown that the technique can provide thorough mechanical information in a wide time-frequency window up to the kHz regime. We have also discussed the practical implications that the insights provided by this technique may have with respect to designing effective strategies for biofilm eradication. Additionally, the technique requires little sample preparation and is compatible with biofilm native environments (e.g., liquids) while providing a high spatial resolution even at the single-cell level. We expect this type of analysis to be also beneficial to other biological (e.g., human cells and tissues) viscoelastic systems that would be better described (compared to the often used Hertzian analysis) by the theoretical framework on which this technique relies.

## V. CONFLICTS OF INTEREST

There are no conflicts to declare.

## VI. ACKNOWLEDGEMENTS

Authors acknowledge support from USDA-NIFA grant 2017-67021-26602.

## VII. REFERENCES

1. J. Otter, K. Vickery, J. d. Walker, E. deLancey Pulcini, P. Stoodley, S. Goldenberg, J. Salkeld, J. Chewins, S. Yezli and J. Edgeworth, *Journal of Hospital Infection*, 2015, **89**, 16-27.
2. C. G. Kumar and S. Anand, *International journal of food microbiology*, 1998, **42**, 9-27.

3. J. Block, K. Haudidier, J. Paquin, J. Miazga and Y. Levi, *Biofouling*, 1993, **6**, 333-343.

4. M. Starkey, M. R. Parsek, K. A. Gray and S. I. Chang, in *Microbial biofilms*, American Society of Microbiology, 2004, pp. 174-191.

5. F. Ahimou, M. J. Semmens, P. J. Novak and G. Haugstad, *Applied and environmental microbiology*, 2007, **73**, 2897-2904.

6. H.-C. Flemming and J. Wingender, *Nature reviews microbiology*, 2010, **8**, 623.

7. B. W. Peterson, Y. He, Y. Ren, A. Zerdoum, M. R. Libera, P. K. Sharma, A.-J. Van Winkelhoff, D. Neut, P. Stoodley and H. C. Van Der Mei, *FEMS microbiology reviews*, 2015, **39**, 234-245.

8. Y. He, B. W. Peterson, M. A. Jongsma, Y. Ren, P. K. Sharma, H. J. Busscher and H. C. van der Mei, *PLoS One*, 2013, **8**, e63750.

9. K. Kovach, M. Davis-Fields, Y. Irie, K. Jain, S. Doorwar, K. Vuong, N. Dhamani, K. Mohanty, A. Touhami and V. D. Gordon, *npj Biofilms and Microbiomes*, 2017, **3**, 1.

10. C.-Y. Hui, T. Tang, Y.-Y. Lin and M. K. Chaudhury, *Langmuir*, 2004, **20**, 6052-6064.

11. A. Gent, *Langmuir*, 1996, **12**, 4492-4496.

12. A. Ahagon and A. Gent, *Journal of Polymer Science: Polymer Physics Edition*, 1975, **13**, 1903-1911.

13. H. Boudarel, J.-D. Mathias, B. Blaysat and M. Grédiac, *npj Biofilms and Microbiomes*, 2018, **4**, 17.

14. K. Kovach, M. Davis-Fields, C. Rodesney and V. Gordon, *cell*, 2015, **16**, 17.

15. A. X. Cartagena-Rivera, W.-H. Wang, R. L. Geahlen and A. Raman, *Scientific reports*, 2015, **5**, 11692.

16. P. D. Garcia, C. R. Guerrero and R. Garcia, *Nanoscale*, 2017, **9**, 12051-12059.

17. C. A. Amo, A. P. Perrino, A. F. Payam and R. Garcia, *ACS nano*, 2017, **11**, 8650-8659.

18. D. J. Müller and Y. F. Dufrene, in *Nanoscience And Technology: A Collection of Reviews from Nature Journals*, World Scientific, 2010, pp. 269-277.

19. D. Martinez-Martin, E. T. Herruzo, C. Dietz, J. Gomez-Herrero and R. Garcia, *Phys Rev Lett*, 2011, **106**, 198101.

20. D. Alsteens, D. J. Müller and Y. F. Dufrêne, *Accounts of chemical research*, 2017, **50**, 924-931.

21. H. Shen, E. A. López-Guerra, R. Zhu, T. Diba, Q. Zheng, S. D. Solares, J. Zara, D. Shuai and Y. Shen, *ACS Applied Materials & Interfaces*, 2018, DOI: DOI: 10.1021/acsami.8b18543.

22. R. Garcıa and R. Perez, *Surface science reports*, 2002, **47**, 197-301.

23. H.-J. Butt, B. Cappella and M. Kappl, *Surface Science Reports*, 2005, **59**, 1-152.

24. C. A. Amo and R. Garcia, *ACS nano*, 2016, **10**, 7117-7124.

25. K. Johnson, 2000.

26. M. Chyasnavichyus, S. L. Young and V. V. Tsukruk, *Langmuir*, 2014, **30**, 10566-10582.

27. P. D. Garcia and R. Garcia, *Biophysical Journal*, 2018, **114**, 2923-2932.

28. D. C. Lin, E. K. Dimitriadis and F. Horkay, *Journal of biomechanical engineering*, 2007, **129**, 904-912.

29. D. C. Lin, E. K. Dimitriadis and F. Horkay, *Journal of biomechanical engineering*, 2007, **129**, 430-440.

30. K. Johnson, K. Kendall and A. Roberts, 1971.

31. D. Maugis, *Journal of colloid and interface science*, 1992, **150**, 243-269.

32. J. L. Hutter and J. Bechhoefer, *Review of Scientific Instruments*, 1993, **64**, 1868-1873.

33. B. V. Derjaguin, V. M. Muller and Y. P. Toporov, *Journal of Colloid and interface science*, 1975, **53**, 314-326.

34. I. N. Sneddon, *International journal of engineering science*, 1965, **3**, 47-57.

35. L. A. Galin, *Contact problems: the legacy of LA Galin*, Springer Science & Business Media, 2008.
36. J. D. Ferry, *Viscoelastic properties of polymers*, John Wiley & Sons, 1980.
37. E. A. López-Guerra and S. D. Solares, *Beilstein Journal of Nanotechnology*, 2017, **8**, 2230-2244.
38. E. A. López-Guerra, F. Banfi, S. D. Solares and G. Ferrini, *Scientific reports*, 2018, **8**, 7534.
39. E. A. López-Guerra, B. Eslami and S. D. Solares, *Journal of Polymer Science Part B: Polymer Physics*, 2017, **55**, 804-813.
40. N. W. Tschoegl, *The phenomenological theory of linear viscoelastic behavior: an introduction*, Springer Science & Business Media, 2012.
41. https://github.com/ealopez.
42. E. Lee and J. R. M. Radok, *Journal of Applied Mechanics*, 1960, **27**, 438-444.
43. G. A. Graham, *International Journal of Engineering Science*, 1965, **3**, 27-46.
44. G. Graham, *Quarterly of Applied Mathematics*, 1968, **26**, 167-174.
45. S. Hunter, *Journal of the Mechanics and Physics of Solids*, 1960, **8**, 219-234.
46. T. Ting, *Journal of Applied Mechanics*, 1966, **33**, 845-854.
47. M. Dao, C. T. Lim and S. Suresh, *Journal of the Mechanics and Physics of Solids*, 2003, **51**, 2259-2280.
48. C. Lim, E. Zhou and S. Quek, *Journal of biomechanics*, 2006, **39**, 195-216.
49. R. Anderssen, A. Davies and F. de Hoog, *Inverse Problems*, 2008, **24**, 035009.
50. F. R. De Hoog and R. S. Anderssen, *Journal of Math-for-Industry*, 2012, **4**, 1-4.
51. Y. M. Efremov, W.-H. Wang, S. D. Hardy, R. L. Geahlen and A. Raman, *Scientific reports*, 2017, **7**, 1541.

52. J. De Sousa, J. Santos, E. Barros, L. Alencar, W. Cruz, M. Ramos and J. Mendes Filho, *Journal of Applied Physics*, 2017, **121**, 034901.

53. F. Mainardi, *Fractional calculus and waves in linear viscoelasticity: an introduction to mathematical models*, World Scientific, 2010.

54. H. Schiessel, R. Metzler, A. Blumen and T. Nonnenmacher, *Journal of physics A: Mathematical and General*, 1995, **28**, 6567.

55. P. D. Garcia and R. Garcia, *Nanoscale*, 2018.

56. E. A. L. Guerra, The George Washington University, 2018.

57. L. Meirovitch, *Fundamentals of vibrations*, Waveland Press, 2010.

58. D. E. Newland, *An introduction to random vibrations, spectral & wavelet analysis*, Courier Corporation, 2012.

59. S. Czesla, T. Molle and J. Schmitt, *Astronomy & Astrophysics*, 2018, **609**, A39.

60. H.-J. Butt and M. Jaschke, *Nanotechnology*, 1995, **6**, 1.

61. R. G. Lyons, *Understanding Digital Signal Processing, 3/E*, Pearson Education India, 2011.